# Understanding and Mitigating Plume Effects During Powered Descents on the Moon and Mars

A White Paper submitted to the Planetary Science Decadal Survey 2023-2032

*Primary Author*
**Ryan N. Watkins**
Planetary Science Institute
rclegg-watkins@psi.edu

*Co-Authors:*
**Philip T. Metzger**, University of Central Florida / Florida Space Institute
**Manish Mehta**, NASA Marshall Space Flight Center
**Daoru Han**, Missouri University of Science and Technology
**Parvathy Prem**, Johns Hopkins University Applied Physics Laboratory
**Laurent Sibille**, NASA Kennedy Space Center / Southeastern University Research Association
**Adrienne Dove**, University of Central Florida
**Bradley Jolliff**, Washington University in St. Louis
**Daniel P. Moriarty III**, Univ. of Maryland College Park / NASA Goddard Space Flight Center
**Donald C. Barker**, MAXD, Inc.
**Ed Patrick**, Southwest Research Institute
**Matthew Kuhns**, Masten Space Systems
**Michael Laine**, LiftPort Group

*Signatories:*
**Michelle Munk**, NASA Langley Research Center
**Andrew Horchler**, Astrobotic
**Robert Mueller**, NASA Kennedy Space Center
**James Mantovani**, NASA Kennedy Space Center
**Nathan Gelino**, NASA Kennedy Space Center
**David Blewett**, Johns Hopkins University Applied Physics Laboratory
**Jeffrey Gillis-Davis**, Washington University in St. Louis

> **Key Recommendation:** All future landed missions on the Moon and Mars must have dedicated measurements of plume-surface interactions, and this data must be made publicly available so we can gain a better understanding of the effects of rocket exhaust on planetary surfaces and can plan for protecting hardware surrounding landing sites.

## 1. INTRODUCTION

During the powered landing of spacecraft on the Moon or Mars, rocket exhaust plumes interact with the surface, altering the physical state of the landing area and creating potential hazards to the spacecraft and nearby hardware. The effects of plume-surface interactions (PSI) differ depending on the planet's gravity, atmospheric density, the mass and thrust of the spacecraft, the physical properties of the soil, and engine configuration. For lunar landings, dust and rocks are blown away at high velocities, spreading particles across the entire surface of the Moon, potentially confounding surface science, and even inserting dust into lunar orbit and impacting orbital hardware (Metzger, 2020). Due to the differing conditions on Mars, a human-class lander would cause a deep crater to form, which in turn would redirect a jet of gas carrying rocks and sand back up at the landing spacecraft. While much has been learned about the physics of PSI over the last few decades, significant gaps in knowledge still exist and can only be filled in by taking measurements during the landing of spacecraft.

With renewed interest in the Moon by the US government, commercial companies, and international entities, and with upcoming Mars surface missions and sample return, understanding exhaust plume interactions with planetary surfaces is crucial to safely land, protect hardware, and conduct scientific investigations. The need is especially true for locations that may require multiple landings at the same site, such as a lunar settlement, and for landing increased masses on both bodies. This white paper provides a broad overview of the current state of knowledge of PSI, potential mitigation strategies, the outstanding questions that remain to be answered, and recommendations for how future missions can contribute to addressing these questions.

## 2. CURRENT STATE OF KNOWLEDGE/OVERVIEW OF PHENOMENOLOGY

### 2.1 Plume Effects on the Moon

On the Moon, PSI for small and mid-sized landers do not generally lead to the creation of a crater beneath the spacecraft, owing to the lack of an atmosphere and to the cohesive and impermeable nature of lunar regolith. Instead, the plume spreads out across the surface and blows material horizontally away from the landing zone at high velocities (Metzger et al., 2011). Plume-surface interactions on the Moon generally begin when the spacecraft is ~30-40 m above the surface. Analysis of digitized Apollo Lunar Module descent videos revealed that plume-lofted dust sheets contained $10^8$-$10^{13}$ particles/m$^3$ and were blown radially away from the descent engines at angles of 0-3° relative to the surface (Immer et al., 2011b). The very fine-grained (dust to sand-sized) materials that compose the regolith surface in the area surrounding the LM were shown to be blown at velocities reaching up to 3 km/s, in some cases exceeding the escape velocity of the Moon (Lane et al., 2008; Metzger et al., 2011). Dust to sand-sized particles in the upper few centimeters of regolith in the area surrounding the lunar module (LM) were blown several kilometers away, leaving the coarser, presumably more compact, underlying soil exposed (Lane et al., 2008). The dust sheet obscures visibility (**Fig. 1**), and in the case of Apollo landers, could cause movement of rocks as large as 10 cm or more (Metzger et al., 2011).

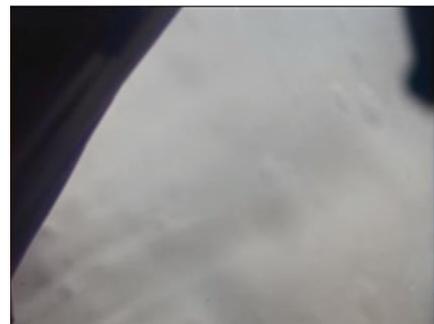

*Fig. 1: Apollo 15 descent video image showing the heaviest episode of dust flow, in which visibility is heavily obscured.*

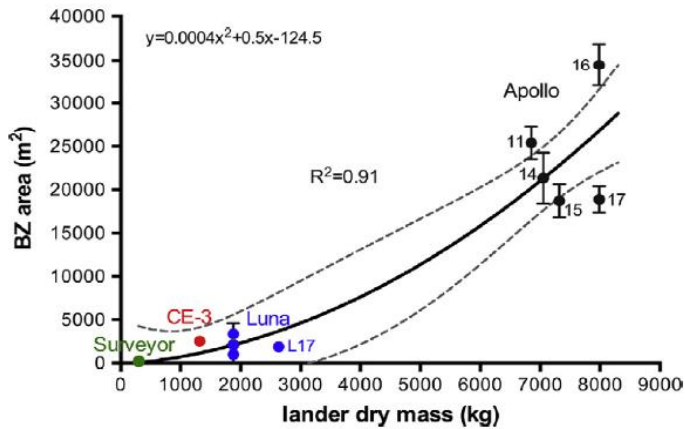

*Fig. 2: The area of the surface visibly disturbed by rocket exhaust increases quadratically with lander mass (Clegg-Watkins et al. 2016).*

**2.2 PSI Scaling with Lander Mass**

At lunar landing sites, visible effects of the surface disturbance extend tens to hundreds of meters out from the lander, depending on lander size (Clegg et al., 2014). Using Lunar Reconnaissance Orbiter (LRO) images, Clegg-Watkins et al. (2016) found a consistent correlation between blast zone area and lander dry mass (**Fig. 2**). While this relationship serves as an important tool in predicting the size of the disturbed areas for future landed missions, it is unknown how variables such as spacecraft design, descent engine configuration, and landing terrain will affect the scale of PSI. Multi-engine configurations could redirect plume spray back up at the spacecraft (Wang, 2019), and hovering could extend the duration of time that PSI occurs. Polar regions with more volatiles and more porous surfaces could experience different surface alterations than non-polar regions. More information and modeling is needed to accurately predict the extent of surface disturbance that will occur, especially by landers much larger than those of Apollo. While blast zones are the visible effects of rocket exhaust interacting with the lunar surface, plume effects go much further than what we observe from orbit.

### 2.1.1. Volatiles from exhaust

Even before a spacecraft reaches the altitudes at which scouring of the surface begins, it releases exhaust gases into the lunar exosphere. Depending on the descent trajectory, some fraction of these gases may escape lunar gravity, but a significant amount may remain gravitationally bound. Understanding the distribution and longevity of exhaust gases in the lunar environment is critical to planning surface operations and interpreting measurements that aim to characterize the extant lunar volatile inventory. This issue is especially important because spacecraft exhaust commonly includes species such as $H_2O$ and $NH_3$ that are among those that exist at the lunar poles (Colaprete et al., 2010).

Recent modeling work (Prem and Hurley, 2019) indicates that a Chang'e 3-class lander may increase the local exospheric density by several orders of magnitude, before exhaust gases are globally dispersed and either cold-trapped (predominantly at the closest pole) or changed by photolysis over the course of a few lunar days. The longevity of exhaust gases in the exosphere depends on both the energetics of gas-surface interactions, as well as the time-scale for photolysis - the main loss process. These processes yield new molecular concentrations in both the exosphere and the regolith after each rocket engine use.

## 2.3 Plume Effects on Mars

Plume-surface interactions on Mars differ from those on the Moon and occur in different regimes, mostly owing to differences in martian regolith, the presence of an atmosphere, and very different thermal environments. Martian regolith is generally less cohesive and more porous, making it more easily excavated by an impinging exhaust plume jet than in the lunar case. The thin atmosphere collimates rocket plumes, as observed for multiple Mars science landers such as Phoenix, InSight, and the Mars Science Laboratory (MSL) Sky Crane. Plume collimation leads to a strong stagnation shock and focused impingement pressures (Mehta et al., 2013), which in turn can lead to large surface erosion and high erosion rates.

On Mars, the dominant regime is bearing capacity failure, which occurs when the exhaust jet places too much load on the soil and creates a crater (Metzger et al., 2009). Owing to these effects, cratering beneath a human-class Mars lander could be mission threatening. Half- and quarter-scale experiments conducted at the Ames Research Center Planetary Aeolian Laboratory with martian simulants at Mars atmospheric pressure showed that the transient crater can reach 1 m diameter in less than 1 second (Mehta et al., 2011a). Another regime was observed on Mars during the Phoenix landing in 2008. Pulsed jets at 10 Hz generated an explosive erosion regime characterized by cyclic fluidization and granular shock waves that propagated through the granular media. This style of erosion leads to the development of large craters and has been shown to produce the highest erosion rates of any mechanism (Mehta et al., 2011b).

## 3. INDUCED HARDWARE AND ENVIRONMENTAL IMPACTS

### 3.1 Hardware Damage

*Surveyor III:* Serendipitously, pieces of Surveyor III provide the best evidence of rocket exhaust plume damage that can occur during a spacecraft landing. Apollo 12 landed ~155 m to the west of Surveyor III and the astronauts returned parts of the spacecraft for analysis on Earth. Studying these pieces has shown that Surveyor III surfaces experienced pitting and cracking as a result of exposure to blowing dust from Apollo 12 (**Fig. 3**). The spacecraft was positioned in a crater slightly below the main spray of dust from the Apollo 12 exhaust plume. Had Surveyor III been exposed to the direct spray, the damage would have been orders of magnitude greater (Immer et al., 2011a). Surveyor III therefore serves as a crucial piece of information in our understanding of how to protect hardware from blowing soil during the descent of spacecraft on the lunar surface.

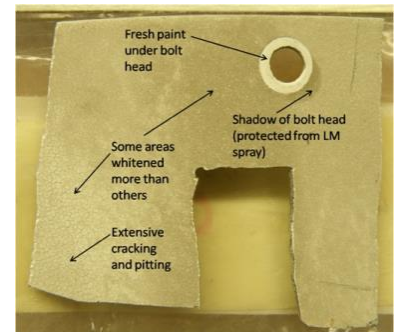

*Fig. 3: Piece of Surveyor III camera shroud with evidence of sandblasting by the Apollo 12 landing (Immer et al., 2011b)*

*Curiosity:* During the landing of MSL in 2012, pebbles thrown up during the rocket-powered landing damaged one of the Curiosity rover's two wind sensors (Gomez-Elvira, 2014). The large size of the Sky Crane descent stage triggered plume-surface interactions ~63 m above the ground, and erosion continuously increased as the vehicle descended (Vizcaino and Mehta, 2015). At ~6 m above the surface, the Sky Crane entered a hovering phase, during which large particles of soil were kicked up, recirculated, and impacted the bottom of the rover chassis. This recirculating of particles likely severed wires on the wind sensor, causing permanent damage.

## 3.2 Operational impacts

The examples of Surveyor III and Curiosity highlight clear operational risks caused by rocket engines operating in proximity of unconsolidated regolith on both the Moon and Mars. Powered descent, hovering, and launch of a spacecraft generate high velocity particles that have the potential to impact a variety of systems mounted on the underside of the spacecraft, or in the region surrounding the launch/landing sites. This hazard requires changes in instrument mounting locations, installation of shields and hardened covers, or deployment mechanisms to configure them for surface viewing post-landing. Risks of impacts increase for multi-engine landers that can cause plume fountaining that directs hot gas and ejecta blast back up towards the vehicle. Uneven cratering/erosion during descent can lead to an increase in slope across the span between lander legs and result in an inclined rest position that may violate launch angle criteria. The same may occur during ascent, resulting in instabilities and/or a damaged zone for the next landing.

Lunar lander designs under development include the capability of decreasing engine thrust or engine shutdowns at a chosen altitude followed by a ballistic landing to minimize PSI. Some configurations include placing landing thrusters near the apex of the lander to avoid using the main engines to complete the descent. The thrust required to launch these vehicles from the lunar surface for a return to orbit would then be provided by main engines close to the surface, thus creating a "worst case" scenario for PSI since the launch will demand near-maximum thrust levels and could blast large amounts of regolith back towards the lander. Apollo ascent vehicles mitigated this risk by using the descent stage as a launch platform that effectively protected the launch vehicle, but little is known of the impact of these launches to the surrounding scientific instruments. Most current lander configurations do not include the abandonment of a descent stage on the surface to enable return to the same landing site.

## 3.3 Environmental impacts and contamination

### 3.3.1 On the Moon

Observations of morphological changes (erosion in clods, surface smoothing) and changes in grain size occurring in the landing zone (Metzger et al., 2011; Kaydash et al., 2011; Clegg et al., 2014), and of the disturbance of surface regolith extending up to hundreds of meters away from the spacecraft, elevate the need for surface mobility by rovers or crew as essential for gathering pristine surface soil samples. Although the Apollo exhaust plumes only excavated regolith to a depth of a few centimeters (Kaydash and Shkuratov, 2012), it is important to consider surface alterations when planning missions that will take samples or study surface features in the immediate vicinity of the lander. The plume also injects volatiles into the local environment, some of which could migrate into cold traps and be measured by instruments that are directed at understanding the volatile distribution in permanently shadowed regions (PSRs) (Shipley et al., 2014). These effects must be quantified so that they can be accounted for in future scientific measurements.

### 3.3.2 On Mars

Experiments performed by Mehta et al. (2011b) established that cyclic shock waves created by pulsed jets propagated through the soil during the Phoenix landing, causing rapid erosion of 5-18 cm depth of soil at rates an order of magnitude higher than other jet-induced processes, and altered the landing surface at radial distances of 20 m. The erosion uncovered subsurface ice (**Fig. 4**) and raised significant concerns about the contamination of the area by ammonia from the plumes of hydrazine engines.

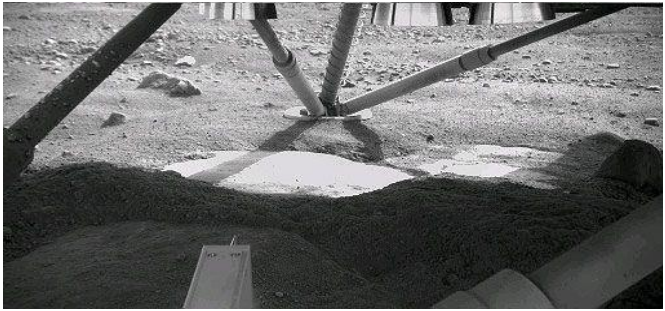
*Fig. 4: PSI during the Phoenix landing exposed subsurface water ice (see Mehta et al., 2011b).*

## 4. MITIGATION TECHNIQUES

Mitigating the adverse effects of PSI is of paramount importance, especially in areas where multiple landings may occur. Below, we briefly outline two potential mitigation techniques, but note that other possibilities exist that we are unable to discuss fully here, and more are rapidly being developed as interests and technologies advance.

### 4.1 Landing/Launch Pads

Because it is difficult to control fluidized soil under a large rocket, one of the best ways to reduce the effects from PSI is to create a solid surface with low erodibility on which a rocket can land. For this reason, landing pads are probably the most important technical and civil engineering structure needed for growth of human activity on the Moon and Mars. Many ideas exist, but few are above a technology readiness level (TRL) of 2 or 3; these include regolith sintering, regolith brick construction and laying, lunar concrete, and alumina injected thruster plumes (Lin et al., 1997; Hintze and Quintana 2013; Mueller et al., 2016, Davis et al., 2017; Meurisse et al., 2018; Fateri et al., 2019; Kuhns, 2020). Inherent limitations of terrestrial testing, and the challenges of the ultra-high vacuum lunar environment, will require the validation of developed techniques on small-scale surface missions to find solutions in the near future.

### 4.2 Propellants

Some propellants may be more suitable than others in situations where contamination of measurements by gases is a concern, e.g., cold-gas thrusters may minimize interference with scientific measurements of lunar volatiles, but also typically generate less thrust. In cases where some contamination is inevitable, we should take inventory of exhaust products and account for their propagation and persistence. Some mass spectrometers set to fly on the early CLPS landers (e.g. the SEAL instrument, Benna et al., 2019) aim to measure spacecraft exhaust, which will be an important step in the development of these models.

## 5. OUTSTANDING QUESTIONS AND STRATEGIC KNOWLEDGE GAPS

Below is a list of outstanding questions that remain regarding plume-surface interactions. This list is not exhaustive, but covers many of the open questions that must be addressed in order to better model PSI, ensure safe landings, and understand physical alterations at the landing sites.

- What is are the size distributions, volumes, and velocities of particles lofted by exhaust plumes? How far do they travel?
- How much dust (mass, particle size) remains lofted after engine shut-down, at what altitude and distribution, and for how long?
- What physical changes occur to the surface in the landing zone?
- What is the nature of contamination of regolith around the landing site? What is the maximum lateral and horizontal range of contamination and how does contamination vary based on different propellants?
- How does PSI differ in different landing terrains?

- Does a fraction of plume gas compounds become part of the lunar atmosphere for extended periods of time, or permanently, and could this affect scientific investigations of the processes that create the lunar exosphere/atmosphere?
- What effects do lander size and engine configuration have on PSI?
- How does the mass/volume/size/velocity/flux of dust particles lofted in the exhaust plume evolve through repeated landings/launches within a zone of interest? Are hazards mitigated or increased with repeated landings/launches?
- How far away from nearby hardware must a landing take place to ensure the safety of the emplaced infrastructure?
- How important will landing pads / prepared surfaces be for repeat missions to an area, and how effective will they be over time?

Answers to these questions would address two open lunar Strategic Knowledge Gaps (SKGs) [https://www.nasa.gov/exploration/library/skg.html] and three open martian SKGs as defined by the Precursor Strategy Analysis Group (P-SAG, 2012). The lunar SKGs will be retired when missions to the lunar surface test transport techniques of lunar resources (**III-A-2**), and are equipped with LIDAR-type instruments to measure the velocity and direction of blast ejecta (**III-D-4**). The martian SKGs (**B4-2, B7-1, B7-2**) will be retired when surface missions are instrumented to take measurements to characterize surface and subsurface regolith properties.

## 6. RECOMMENDATIONS

Much is still not understood about the effects of rocket exhaust on the lunar and martian surfaces during the powered descent and ascent of spacecraft. In order to address outstanding questions and SKGs regarding lunar and martian PSI, and to plan for protecting hardware surrounding landing sites, **all future landed missions on the Moon and Mars must have dedicated measurements of plume effects, and this data must be made publicly accessible.** *Note: These recommendations are for NASA spacecraft/instruments; separate recommendations could be established for commercial (e.g., CLPS) providers and are beyond the scope of this white paper.*

**Recommendation #1: Include dedicated descent imagers on every surface mission so that PSI can be directly recorded and reviewed by ground teams.** The primary, and easiest, method of obtaining PSI data is to have dedicated descent imaging systems operating during landing, providing quantifiable information that can be correlated with known camera and spacecraft parameters (altitude, engine configuration, camera angles, etc.). This could include dedicated images of the surface beneath and surrounding the lander to record morphological changes.

Several imaging systems are currently in development to fly on upcoming CLPS missions. The Stereo Cameras for Lunar Plume Surface Studies (SCALPSS) (Munk, 2019) and the Heimdall imaging systems (Yingst et al., 2020) will record observations of PSI effects during descent and will image the surface beneath the lander after touchdown.

**Recommendation #2**: **As far as possible, make all data related to PSI effects publicly accessible** for the community at large to use for planning future missions, improving plume effects models, and developing instruments for future studies.

**Recommendation #3: Develop methods and instruments for making key measurements of PSI;** *i.e.,* **the size and velocity of ejecta particles.** These measurements are needed for validating and improving PSI models and can be done with a variety of instruments and techniques, including, but not limited to: cameras, lasers, mid-wave infrared sensors, and dust collectors/sticky pads.

The Aerogel Experiment Gathering Impactor Samples (AEGIS) concept, under development at NASA Goddard, utilizes aerogel panes to capture particles blown during landings (Moriarty et al., 2020). Deployed at the lunar surface by a CLPS lander or Artemis astronaut, they would then be returned (and possibly replaced) by later missions for analysis in terrestrial laboratories.

**Recommendation #4**: **Assess and record key flight data**, including the heating and pressure environments on the base of the lander since these can affect the materials, design, and mass of the vehicle. Record detailed information on camera angles and altitude during descent. These data will be crucial for interpreting and simulating results (e.g., Munk, 2019).

**Recommendation #5**: **Invest funding into studies of long-term infrastructure architectures and mitigation techniques**, including landing pads, berms, propellant variations, and new types of dust sensors. Devise plans to implement infrastructure development as soon as possible.